\documentclass[sigconf]{acmart}

\AtBeginDocument{%
  \providecommand\BibTeX{{%
    \normalfont B\kern-0.5em{\scshape i\kern-0.25em b}\kern-0.8em\TeX}}}

\copyrightyear{2021}
\acmYear{2021}
\setcopyright{acmcopyright}
\acmConference[WUWNet'21]{The 15th International Conference on Underwater Networks & Systems}{November 22--24, 2021}{Shenzhen, Guangdong, China}
\acmBooktitle{The 15th International Conference on Underwater Networks \& Systems (WUWNet'21), November 22--24, 2021, Shenzhen, Guangdong, China}
\acmPrice{15.00}
\acmDOI{10.1145/3491315.3491323}
\acmISBN{978-1-4503-9562-5/21/11}

\acmSubmissionID{108193}

\begin{document} \sloppy

%%
%% The "title" command has an optional parameter,
%% allowing the author to define a "short title" to be used in page headers.
\title{Underwater Acoustic Communication Channel Modeling using Deep Learning}

%%
%% The "author" command and its associated commands are used to define
%% the authors and their affiliations.
%% Of note is the shared affiliation of the first two authors, and the
%% "authornote" and "authornotemark" commands
%% used to denote shared contribution to the research.
\author{Oluwaseyi Onasami}
\authornote{Main Author}
\email{oonasami@pvamu.edu}
\orcid{1234-5678-9012}
\affiliation{%
  \institution{Center of Excellence in Research and Education for Big Military Data Intelligence, Prairie View A\&M University}
  \city{Prairie View, TX 77446}
  \country{USA}}

\author{Damilola Adesina}
\authornote{Contributing Authors}
\email{dadesina@pvamu.edu}
\affiliation{%
  \institution{Center of Excellence in Research and Education for Big Military Data Intelligence, Prairie View A\&M University}
  \city{Prairie View, TX 77446}
  \country{USA}}

\author{Lijun Qian}
\authornotemark[2]
\email{liqian@pvamu.edu}
\affiliation{%
  \institution{Center of Excellence in Research and Education for Big Military Data Intelligence , Prairie View A\&M University}
  \city{Prairie View, TX 77446}
  \country{USA}}

%%
%% By default, the full list of authors will be used in the page
%% headers. Often, this list is too long, and will overlap
%% other information printed in the page headers. This command allows
%% the author to define a more concise list
%% of authors' names for this purpose.
\renewcommand{\shortauthors}{Oluwaseyi O, et al.}

%%
%% The abstract is a short summary of the work to be presented in the
%% article.
\begin{abstract} 
With the recent increase in the number of underwater activities, having effective underwater communication systems has become increasingly important. Underwater acoustic communication has been widely used but greatly impaired due to the complicated nature of the underwater environment. In a bid to better understand the underwater acoustic channel so as to help in the design and improvement of underwater communication systems, attempts have been made to model the underwater acoustic channel using  mathematical equations and approximations under some assumptions. In this paper, we explore the capability of machine learning and deep learning methods to learn and accurately model the underwater acoustic channel using real underwater data collected from a water tank with disturbance and from lake Tahoe. Specifically, Deep Neural Network (DNN) and Long Short Term Memory (LSTM) are applied to model the underwater acoustic channel. Experimental results show that these models are able to model the underwater acoustic communication channel well and that deep learning models, especially LSTM are better models in terms of mean absolute percentage error. 
\end{abstract}

%%
%% Keywords. The author(s) should pick words that accurately describe
%% the work being presented. Separate the keywords with commas.

%% This command processes the author and affiliation and title
%% information and builds the first part of the formatted document.
\maketitle

\section{Introduction}
\label{sec:introduction}

In recent years, there seem to be a growing interest in researches bothering around underwater wireless communication by both civil and military entities. This is largely due to the surge in the number of underwater activities such as the underwater surveillance by the military, underwater mining, laying of pipeline and fibre optic cables, aquatic/biological research and documentaries. About 71\% of the earth is covered with water thus there is a need by marine biologists, engineers and other concerned researchers for technologies to explore the underwater environment even more~\cite{MACUWCS2012}. Due to the increasing underwater activities, there is the crucial need for an efficient underwater communication systems. 

The underwater environment is complicated and recognized as one of the most complex communication medium~\cite{MAUACCB2017}. The underwater acoustic (UWA) communication medium, when compared to terrestrial radio systems,  limits effective communication due to  its extremely slow propagation , low available bandwidth , large multipath delay spread , etc~\cite{chanmodelUAN2020}, thus making modeling of the underwater acoustic channel quite difficult~\cite{chanmodelIOT2021}.

There are two major established methods of transmission for underwater wireless communication. These methods are transmission through acoustic and electromagnetic media. Transmission through acoustic medium involves the use of acoustic waves which are as a result of the physical vibrations of particles while transmission through electromagnetic medium involves the use of electromagnetic waves. Electromagnetic waves are the result of the interference of the electric and magnetic fields. Both can be deployed for underwater communication. However, owing to the physical nature of these waves, the acoustic waves perform better in the underwater environment. The demerits of electromagnetic waves are the high power consumption, large size of the antenna~\cite{MACUWCS2012} and the inability to propagate over long distances in an underwater environment except at extremely low frequencies. Transmitting signals at such low frequencies require expensive powerful transmitter which makes the use of electromagnetic waves an expensive option~\cite{ChitreHighFreq2007}.

\begin{figure}[] 
	 \centering
    	 \includegraphics[width=\linewidth]{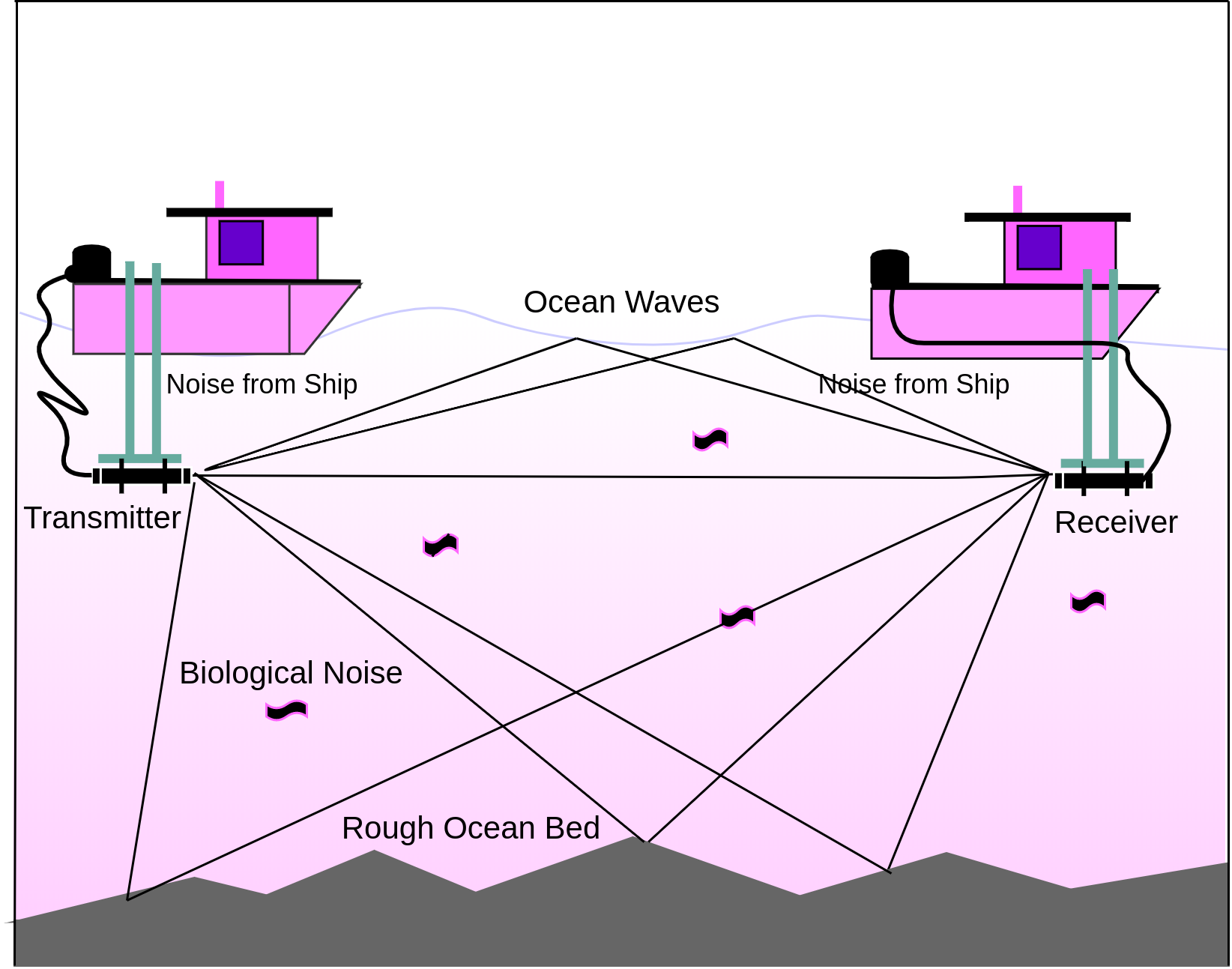}
      	\caption{Underwater Communication Scenario~\cite{LuoCognitive2014}}
    \label{fig:underwater_image}
\end{figure}

\begin{figure*}[htbp]
	 \centering
    	 \includegraphics[width=\linewidth]{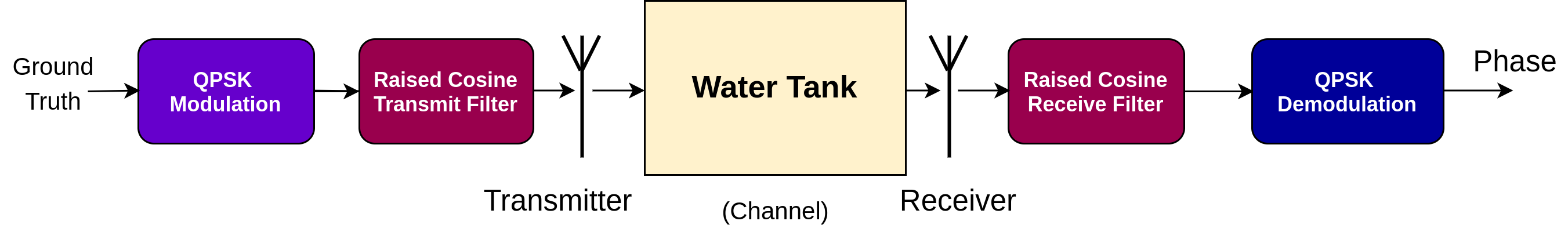}
      	\caption{Data 1 Generation and Collection Pipeline}
    \label{fig:data_1}
\end{figure*}

On the contrary, the power requirement for acoustic wave propagation is smaller, making acoustic transmission preferable to most underwater researchers. However, it is noticed that acoustic wave becomes impractical when it comes to communication involving water-air interface while the electromagnetic wave can be deployed. 
UWA communication has  application in areas including, but not limited to, the off-shore oil industry, marine commercial operations, defense, and oceanography. The propagation of wireless acoustic signals through a water body as the channel of propagation is greatly affected by the marine or underwater environment. The UWA channel suffers a lot of obstacles such as the Doppler shift, strong multi-path propagation and high attenuation, limited bandwidth, severe fading, long delay spread, rapid time variation of channel, path loss and noise~\cite{StojanovicUWA2009, chanmodelUWCN2008}. As a result, researching and having a good understanding of the effect of the underwater environment on the communication signal becomes paramount in order to design and develop effective underwater communication systems. An effective approach is to appropriately model the underwater acoustic channel by simulating the effect of real water environment parameters on the channel~\cite{MAUACCB2017}. Figure~\ref{fig:underwater_image} depicts a typical underwater acoustic communication scenario. Variation in sound velocity, roughness of the ocean bed, multi-path propagation of acoustic signals and ambient ocean acoustic noises caused by aquatic animals and human activities make channel modeling very challenging~\cite{StojanovicUWA2009}. 

Some mathematical underwater acoustic channel models have been developed~\cite{StojanovicUWA2009}.  A commonly used one for channel modeling in UWA communications is the BELLHOP model which is an open source beam/ray tracing model for estimating acoustic pressure fields in the underwater environment. According to~\cite{chanmodelUAN2020}, BELLHOP has a number of modifications that let researchers use it in their research, such as VirTEX for modeling time-varying UWA channels and the World Ocean Simulation System (WOSS)  for modeling underwater acoustic network (UAN) in an environment that represents a specific geographical area. However, most of these models are based on mathematical assumptions and approximations and not built with real underwater communication data~\cite{MAUACCB2017}.

The application of machine learning, or artificial intelligence at large, has recorded great successes in areas like image \& voice recognition, language processing, medical diagnosis and wireless communications. This is because of its ability to learn and intelligently respond to dynamic and complex operating conditions such as we have in UWA communication channel. It is also believed that the knowledge of current operating conditions and environment leveraging on big data analytics~\cite{QianJCIN2017} will allow communication systems to make the best opportunistic decisions. However, not much work has been done in the area of underwater acoustic communications. This is largely due to the complicated nature of the underwater environment and the unavailability of sufficient and good data. Consequently, in this work, we leverage on the capability of the machine learning to build an underwater acoustic channel model and validate with field experiments using real underwater data. \\ \\ The contributions of this paper are:
\begin{enumerate}
    \item In order to mitigate the unrealistic assumptions made by mathematical models for underwater acoustic communications and take advantage of the emerging experimental data, a data driven approach is proposed for underwater acoustic channel modeling in this work.
    \item Underwater acoustic channel modeling using traditional machine learning and deep learning methods have been carried out. Observations and insights are provided based on the comparison of the results from various models.
\end{enumerate}

The remainder of this paper is structured as follows: Section~\ref{sec:datas} describes the data generation and collection pipeline and gives the description of the datasets used in this work. Section~\ref{sec:modelused} highlights the deep learning and traditional machine learning algorithms used in this experiment. An overview of the result is given in Section~\ref{sec:results} and observations from the result are discussed. Some related works in the recent past are shown in Section~\ref{sec:relatedwork} while Section~\ref{sec:conclusions} concludes the paper.

\section{Data Generation and Description}
\label{sec:datas}
For this experiment, three different datasets were used to evaluate the performance of traditional machine learning and deep learning  models in modeling the underwater acoustic channel. The datasets are described as follows:

The first datasets, subsequently referred to as \textit{Data 1}, were obtained from a developed test bed. The test bed was built using water tank with no disturbance. The acoustic source (the transmitter) and the receiver were placed horizontally apart and at a perpendicular distance below the water level. Figure~\ref{fig:data_1} shows the general pipeline for \textit{Data 1} generation and collection. Digital message signals were first passed through a quadrature phase shift keying (QPSK) modulation block and outputs continuous signals. The continuous signals were then passed through a raised cosine transmit filter. The filtered-QPSK modulated continuous signals were then sent over the underwater channel through the acoustic transmitter.. These continuous signals were then captured by the sonar at the receiving end immediately after the channel. The sample rate of acquisition is $1,000,000$ and the length of each data object is $60$ seconds. The sonar working frequency is $200$kHz and the real digital signal transfer speed is $2K/s$. The data collected at the transmitter, just before the channel were used as the input to the machine/deep learning models while the signals collected at the receiver, immediately after the channel were used as the label for training the models.
 
 \begin{figure}[h] %htbp
	 \centering
    	 \includegraphics[width=\linewidth]{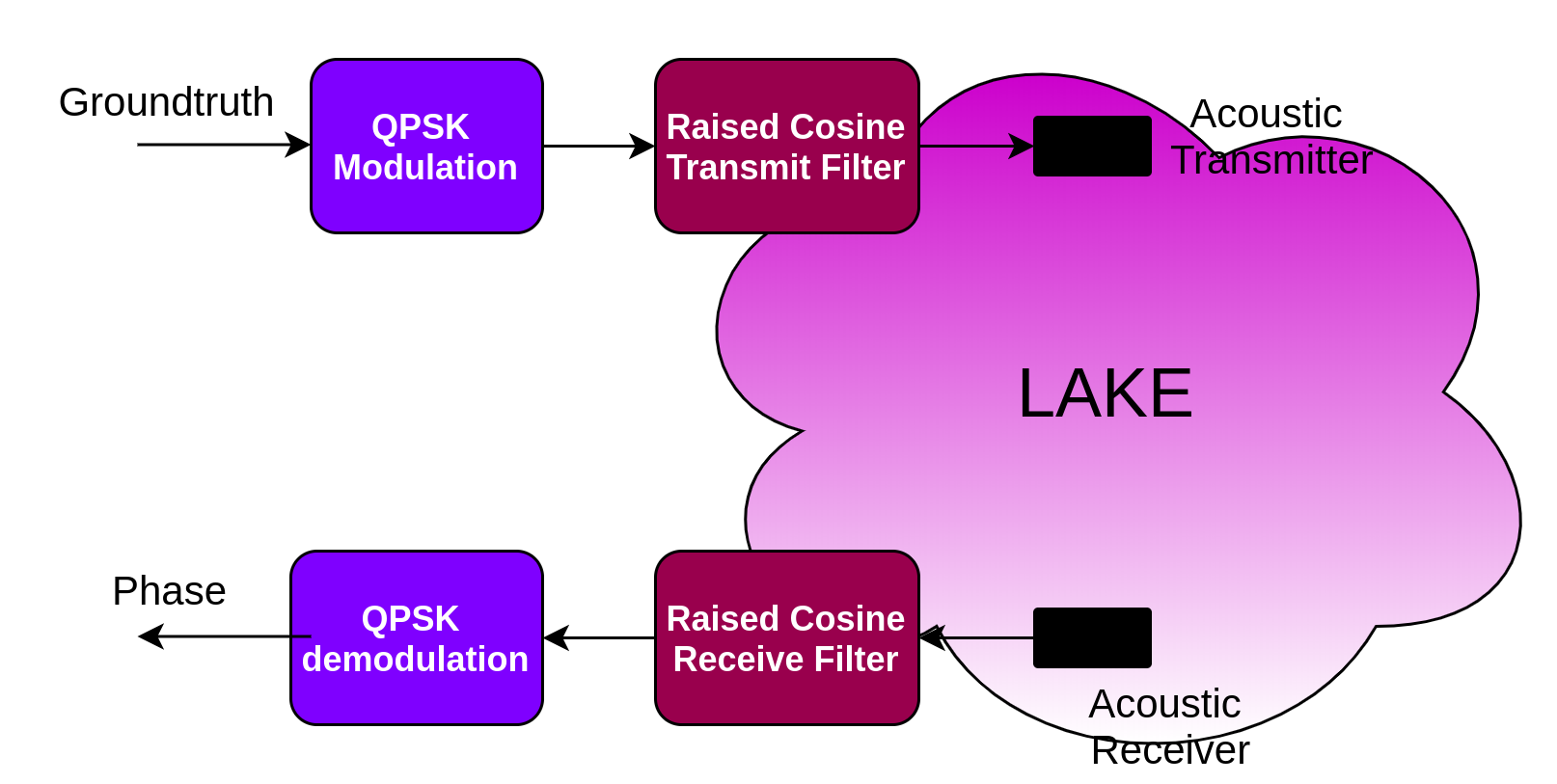}
      	\caption{Data 2 Generation and Collection Pipeline}
    \label{fig:data_2}
\end{figure}

The second category of data, hereafter referred to as \textit{Data 2} were collected from a lake, a natural body of water surrounded by land, also with no artificial disturbance introduced. The transmitted symbols were subjected to the same signal processing stages as with \textit{Data 1}  and sent over the natural water body to have a real underwater dataset with the same sonar working frequency, sampling rate, transfer speed, horizontal distance between the transmitter and the receiver and perpendicular distance into the water, as illustrated in Figure~\ref{fig:data_2}. In the same manner, data at the transmitter and those taken immediately after the channel at the receiver were used as the input and label to the machine learning models respectively. The third category of data, hereafter referred to as \textit{Data 3}, were collected using the same setup and parameters for the lake data, however with disturbance to create a mild chaotic scenario. For the three categories of data, $60,000,000$ samples were collected. 

\section{Machine Learning Models}
\label{sec:modelused}
Machine learning models have proven, in the last decade, to be successful in many areas of application such as image recognition, times series forecasting and sentiment analysis. As mentioned in Section~\ref{sec:introduction}, the underwater acoustic data is a continuous time-series data; one-dimensional in space. Modeling the underwater acoustic communication channel is a regression task. The input to the underwater channel is a sequence of data while the output of the channel is also a sequence of data of the same size received at the receiver. However, due to the nature of these data, not all regression models can be deployed. Models to be considered must be able to take in sequence of data and also output sequence of data as their predictions. In~\cite{ECMLTS2010}, an extensive comparison study of the some machine learning models for time series data was carried out. The study highlighted some machine learning models that can be used for time series data and also established that time series preprocessing methods have shown to have impacts on the performance of the models. 

For a relative evaluation of these machine learning models, we used two established deep learning models - Deep Neural Network (DNN) and the Long Short Term Memory (LSTM)  which is  a special kind of RNN. Deep learning is a branch of machine learning that uses multiple hidden layers in a neural network. It models functions of increasing complexity by adding more layers and more non-linear processing neurons within a layer. It also has ability to learn higher level representations of input data~\cite{LeCunNature2015}.

\subsection{Traditional Machine Learning}
In this study, we used some traditional machine learning models that met the criteria described above. These models include the k-Nearest Neighbour, Random Forest, Linear Regressor and the Multi-Layer Perceptron. 
\subsubsection{K Nearest Neighbor:}
The k-nearest neighbors (KNN) is a supervised machine learning algorithm technique that may be used for both classification and regression issues.  The algorithm believes that objects that are similar are close together, that is, related items are close together. The algorithm hinges on this assumption to capture the similarity between objects by calculating the distance between the objects or points. This algorithm is straightforward and simple to implement as it does not require creating a model nor tuning many hyper-parameters, the major hyper-parameter being the number of neighbors, $K$. We ran the KNN algorithm numerous times with different values of $K$ to find the $K$ that decreases the amount of errors we encounter while keeping the algorithm's capacity to properly make predictions.
\subsubsection{Random Forest:}
A random forest is a supervised machine learning approach based on decision tree algorithms.  It makes use of ensemble learning, which is a technique for solving complicated problems by combining several simpler models. The random forest is made up of many decision trees and the resulting forest is trained either by bagging or bootstrap aggregation. The algorithm determines the outcome based on the decision trees' predictions and the final prediction is based on  averaging the output of various trees. The performance of the predictions  improves as the number of trees increases. It can also be applied to address problems involving regression and classification. Without hyper-parameter adjustment, random forest can provide a reasonable prediction and it also overcomes the problem of overfitting in decision trees.
\subsubsection{Linear Regression:}
Linear regression analysis is a statistical technique for predicting the value of one variable based on the value of another. The dependent variable is the variable you want to predict while the independent variable is the one you're using to forecast the value of the other variable.
This type of analysis can involve one or more independent variables that best predict the value of the dependent variable in order to estimate the coefficients of the linear equation. Linear regression creates a straight line that reduces the difference between expected and the groundtruth. Some linear regressors employ the least-square  approach to determine  the line of best-fit for a collection of available.
\subsubsection{Multi-Layer Perceptron:}
The multi layer perceptron (MLP) is a feed forward neural network augmentation consisting of the input layer, hidden layers and the output layer.  The input signal to be processed is passed through the input layer. The output layer is responsible for tasks such as prediction and categorization depending on if it is regression or classification problem. The inputs are pushed forward through the MLP in the same way that they are in the perceptron by taking the dot product of the input with the weights that exist between the input layer and the hidden layer. At the hidden layer, this dot product returns a value. The MLP then applies activation functions at each of the hidden layers and pushes the calculated output at the current layer to the next layer in the MLP by taking the dot product with the corresponding weights after it has been pushed through the activation function. The calculations will be employed in the output layer for either a backpropagation method that corresponds to the activation function chosen for the MLP  for training or a decision will be made based on the output when testing.

\subsection{Deep Neural Network}
\label{sec:dnn}
\begin{figure}[htbp]
	 \centering
    	 \includegraphics[width=\linewidth]{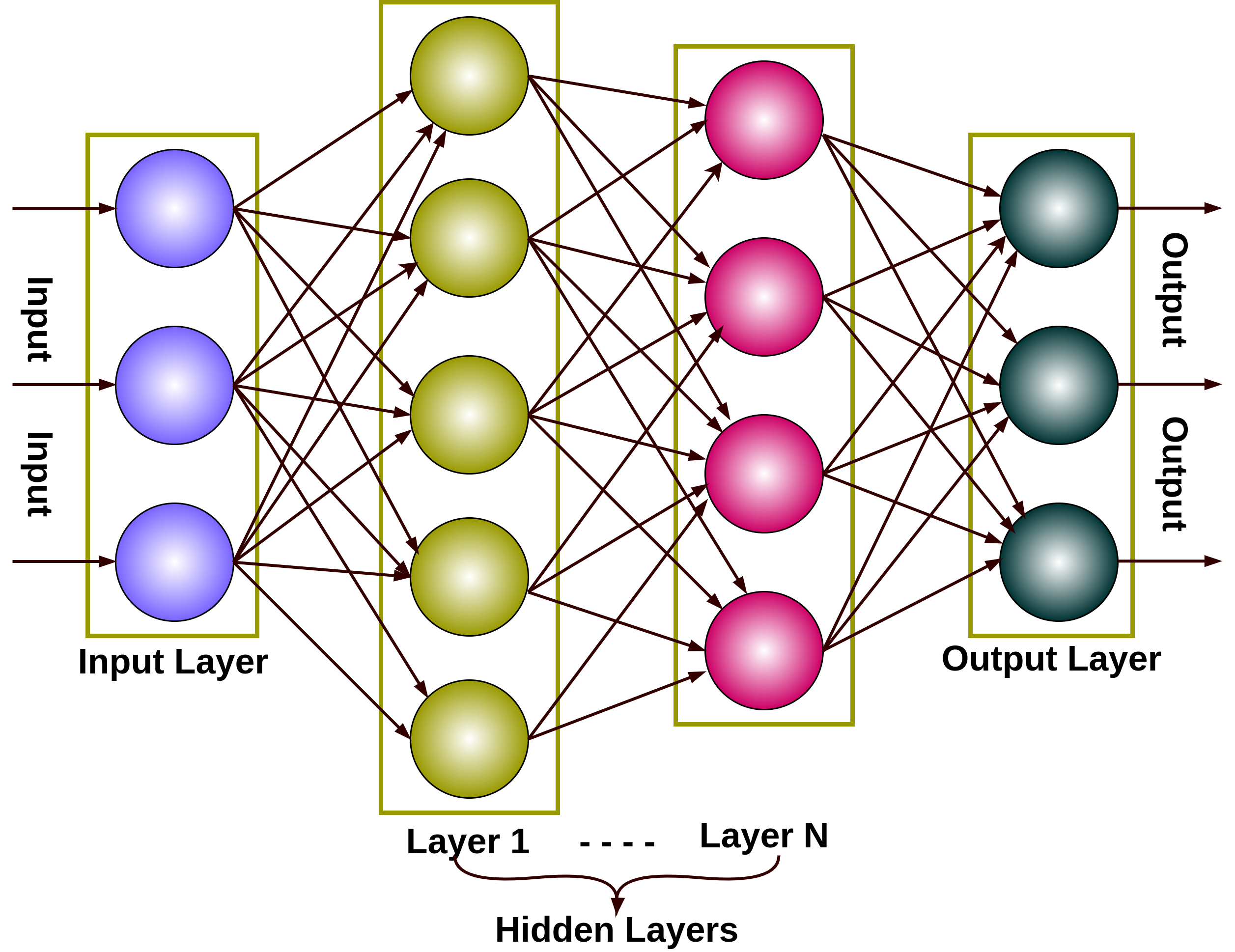}
      	\caption{A Deep Neural Network with N hidden layers }
    \label{fig:dnn}
\end{figure}

A deep neural network (DNN), typically, consist of more than one hidden layer that are fully connected ~\cite{BengioNOW2009, GoodBengCour2016} as shown in Figure~\ref{fig:dnn}. Each hidden layer has several nodes which are termed the hidden units. For this experiment, we used two different DNN architecture. The first architecture contains four(4) dense layers. Each dense layer has $256$ nodes except the output dense layer which has $578$ nodes corresponding to the number of samples corresponding to one transmitted symbol. The first three layers use the Rectified Linear Unit (ReLU) activation functions while the output layer has no activation function. 

\begin{figure}[htbp]
	 \centering
    	 \includegraphics[width=\linewidth]{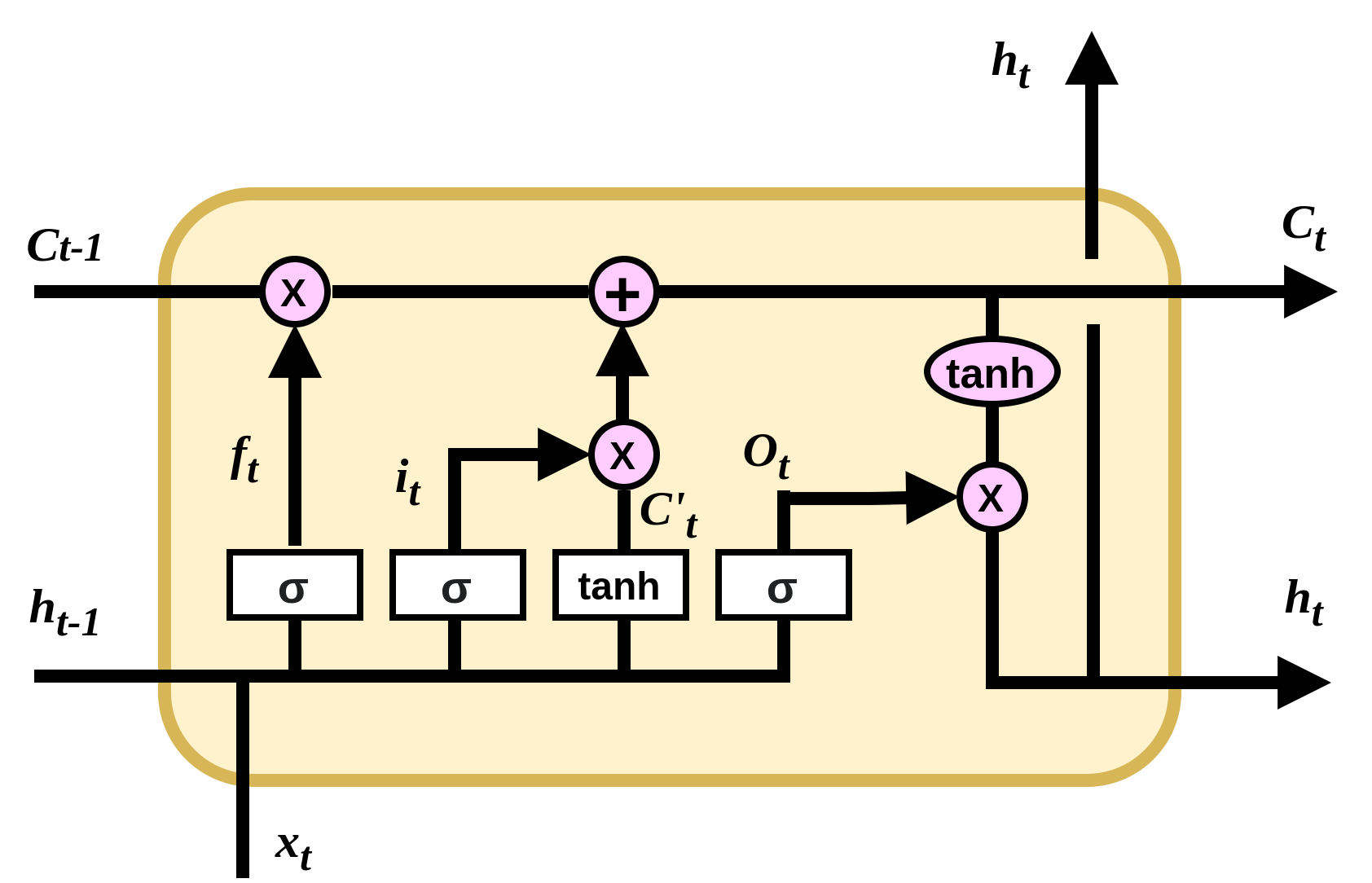}
      	\caption{Long Short Term Memory~\cite{applstm2019}}
    \label{fig:lstm}
\end{figure}

For the second architecture, we only increased the number of the hidden layers to six(6) to make the model deeper and improve the performance of the model on the dataset with disturbance. These models were developed using Keras framework. The datasets were reshaped into number of training examples ($N_X$) times the number of samples per symbol ($N_S$); a two-dimensional tensor appropriate for the Keras dense layer, to form input shape into the network. For the DNN and LSTM in Section~\ref{sec:lstm}, the mean squared error loss function and Adam optimizer were used.  

\subsection{Long Short Term Memory}
\label{sec:lstm}
The Long Short Term Memory (LSTM) is a variant of the  Recurrent Neural Network. RNNs are used generally for tasks that require sequential inputs, such as speech and language. The RNN gets input sequence one symbol at a time and this is held in their hidden units. These hidden units holds information relating to the history of all the past input of the sequence. The architecture of an LSTM unit is as shown in Figure~\ref{fig:lstm}.

\begin{equation}
\begin{aligned}
    f_t = \sigma(W_f \cdot [h_{t-1}, x_t] + b_f) \\
    i_t = \sigma(W_f \cdot [h_{t-1}, x_t] + b_i) \\
    o_t = \sigma(W_f \cdot [h_{t-1}, x_t] + b_o) \\
    c^\prime_t = \tanh(W_c \cdot [h_{t-1}, x_t] + b_C) \\
    c_t = f_t \ast c_{t-1} + i_t \ast c^\prime_t \\
    h_t = o_t \ast \tanh(c_t) 
    \end{aligned}
\end{equation} where $b_f, b_i, b_C$ and $b_o$ are biases, $x_t$ is the previous layer or input.
The LSTM has the ability to learn input data with long-term dependencies and this makes it preferable because traditional RNN performance declines when the distance between the relevant information and the point where it is needed becomes very large~\cite{HochreiterACM1997}. It has the advantage of learning and remembering over long sequences.

In a similar manner, two different LSTM architectures were used in this experiment. The first architecture is made up of two LSTM layers and a dense layer as the output layer. For the second architecture, there were six(6) LSTM layers and one dense layer. Each LSTM layer has $320$ memory blocks with no activation function. The dense layer which doubles as the output layer also has $578$ nodes. The data was reshaped to fit the LSTM layer expected input of three dimensional tensor appropriate for Keras LSTM layer. Thus, the input data to the network is a $N_X$ $\times$ $N_{T}$ $\times$ $N_S$ tensor, where $N_T$ is the number of time steps set to window size which defines the number of input variable used to predict the next time step. 
\section{Results and Discussion}
\label{sec:results}
Table~\ref{machine} shows the test results for \textit{Data 1} and \textit{Data 2} using traditional machine learning methods and deep learning methods. The model performances were evaluated using the mean absolute percentage error (MAPE). It is the average of the absolute percentage errors of predictions. Percentage errors are summed without regard to sign to compute the MAPE as seen in equation (\ref{eqn:seven}).
\begin{equation}
   MAPE = \frac{\mathrm{1}}{n}  \sum_{t=1}^{n} \left| \frac{{A_t - F_t}}{A_t} \right|
   \label{eqn:seven}
\end{equation} where $A_t$ is the actual value and $F_t$ is the predicted value.

\begin{table}  [] %\scriptsize
\caption{Performance of machine learning models on Data 1 and Data 2 (with no disturbance)}
\label{machine}
\centering
\begin{tabular}{llcccc}
\toprule
\multicolumn{2}{c}{}& \textbf{Data 1} & \textbf{Data 2} \\
 \midrule
\textbf{Model} & \textbf{Data size} & \textbf{MAPE(\%)} & \textbf{MAPE(\%)}\\ 
\midrule
Multi-layer Perceptron& 100,000$\times$578 & 5.56 & 11.58 \\ 
K-Nearest Neighbor& 100,000$\times$578 &0.87 &9.54\\ 
Linear Regressor& 100,000$\times$578 &1.80 &12.24 \\ 
Random Forest& 100,000$\times$578&  1.64&10.95 \\ 
Deep Neural Network& 100,000$\times$578&  2.04&2.22\\ 
LSTM & 100,000$\times$578&  1.46&2.21  \\ 
\bottomrule
\end{tabular}
\end{table}

\begin{table} []  %\scriptsize
\caption{Performance of the deep learning models on Data 3 (with disturbance) using the same deep learning architecture as Table~\ref{machine}}
\label{deeper}
\centering
\begin{tabular}{llccc}
\toprule
\multicolumn{3}{c}{\textbf{Real Data}} \\
\midrule
\textbf{Model} & \textbf{Data size} & \textbf{MAPE(\%)} \\ 
\midrule
Deep Neural Network & 100,000$\times$578 & 5.17\\ 
LSTM Network & 100,000$\times$578  & 3.24\\ 
\bottomrule
\end{tabular}
\end{table}

\begin{table} [] %\scriptsize
\caption{Performance of the deep learning models with deeper architecture on Data 3 (with disturbance)}
\label{deep}
\centering
\begin{tabular}{llccc}
\toprule
\multicolumn{3}{c}{\textbf{Real Data}} \\
\midrule
\textbf{Model} & \textbf{Data size} & \textbf{MAPE(\%)} \\ 
\midrule
Deep Neural Network & 100,000$\times$578 & 4.19\\ 
LSTM Network & 100,000$\times$578  & 3.14\\
\bottomrule
\end{tabular}
\end{table}

\begin{figure}[]
	 \centering
    	 \includegraphics[width=\linewidth]{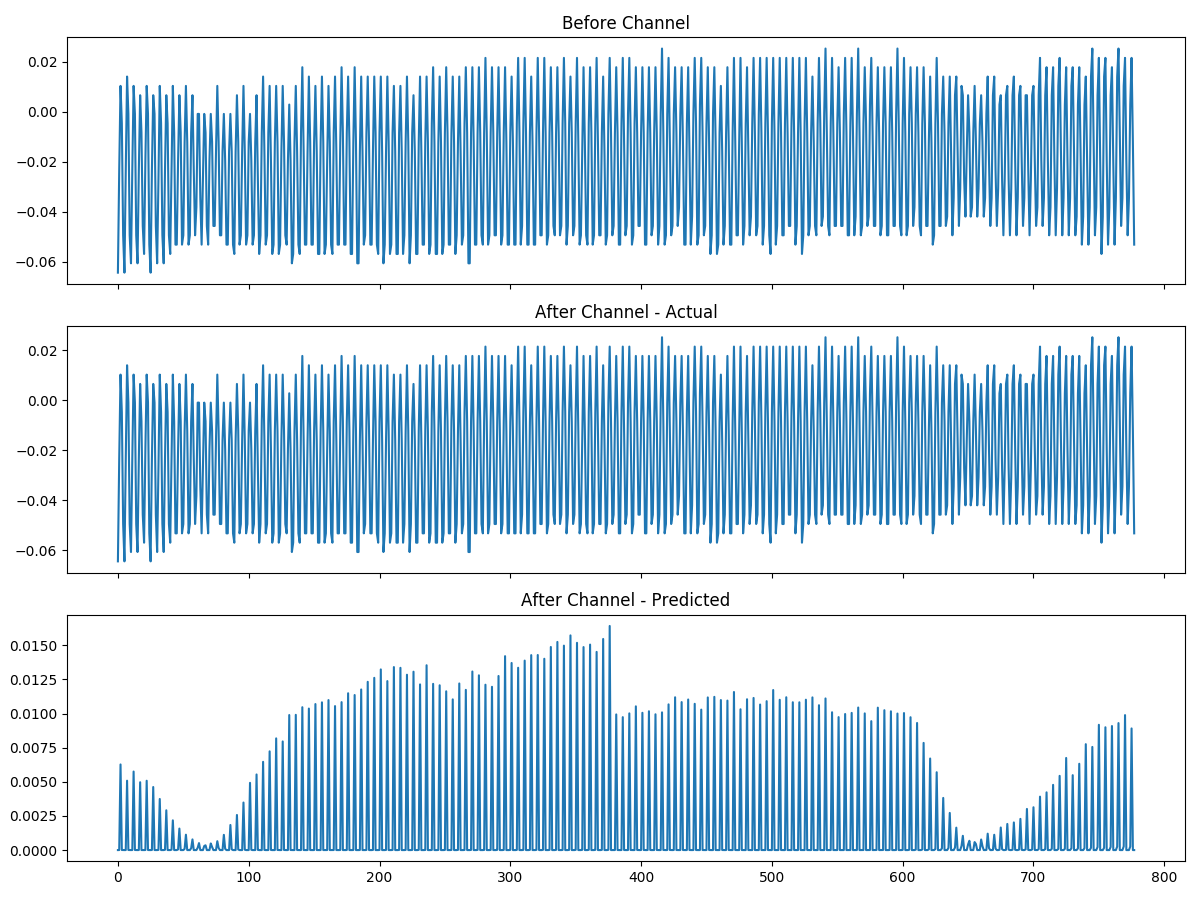}
      	\caption{Sample Plot of Linear Regression Predicted Data with Data 2 }
    \label{fig:linear}
\end{figure}

\begin{figure}[]
	 \centering
    	 \includegraphics[width=\linewidth]{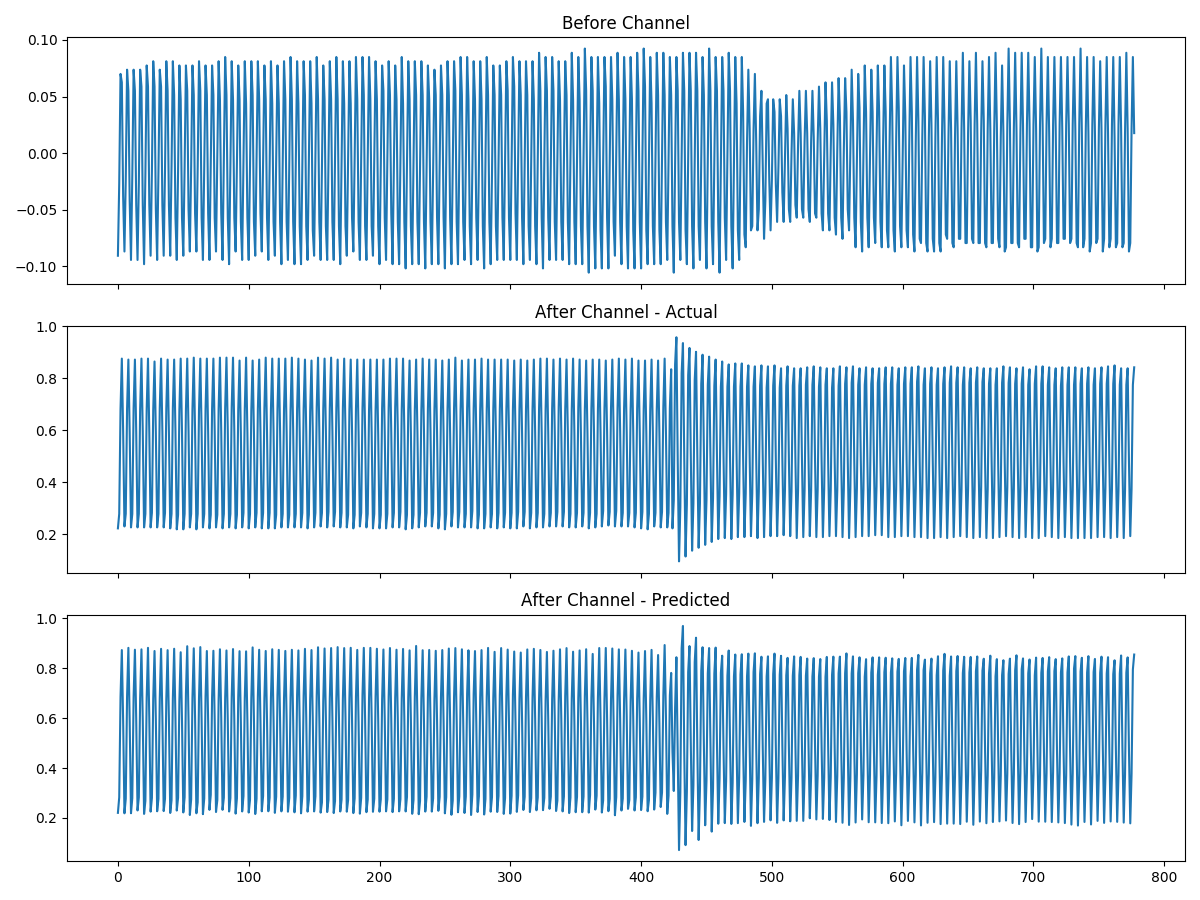}
      	\caption{Sample Plot of an LSTM Predicted Data with Data 3}
    \label{fig:lstm}
\end{figure}

In evaluating regression problems, MAPE gives very intuitive interpretation in terms of relative error and is preferred in the evaluation because it provides the error in terms of percentages and the problem of positive and negative errors canceling each other out is avoided. The smaller the MAPE the better the prediction.

From the experimental results in Table~\ref{machine}, it is clear that the performance of the model became worse with real underwater data, even with no artificial disturbance introduced compared to the data collected from water tank. Different machine learning models differ in their performances. Of the traditional machine learning models under consideration, k-Nearest Neighbors (kNN) has the best performance across the two datasets (Data 1 and Data 2) while the multi-layer perceptron and linear regressor have the worst performances. Figure~\ref{fig:linear} shows a sample predicted result using the linear regressor. It is observed that the predicted results (third plot in Figure~\ref{fig:linear}) are far from the ground truth (second plot in Figure~\ref{fig:linear}), and  linear regressor could not model the underwater acoustic channel well.

On the contrary, the deep learning models - LSTM and DNN - did better with real underwater data and thus were considered for use with \textit{Data 3} with disturbance. For both the LSTM and DNN models, we used the learning rate of $0.001$, batch size of $64$ and epoch of $100$. 

\begin{figure}[h]
	 \centering
    	 \includegraphics[width=\linewidth]{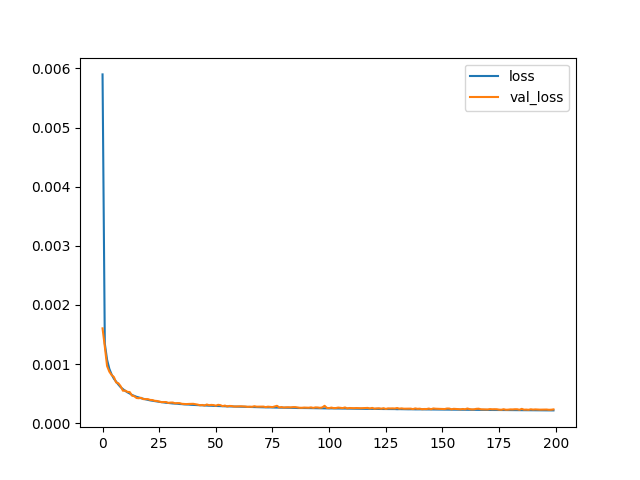}
      	\caption{Loss Curve for the LSTM model}
    \label{fig:loss_lstm}
\end{figure}

When the LSTM and DNN models were trained with $Data\ 3$, with the same architecture and hyperparameters, their performances became worse as seen in Table~\ref{deeper}. Recall that $Data\ 3$ is more chaotic because of the external disturbance introduced to the lake. This prompted the use of deep models with deeper architectures as described in Section~\ref{sec:modelused}, and the LSTM and DNN models gave a better performance as shown in Table~\ref{deep}. 
It is also observed in all cases, $Data\ 1$, $Data\ 2$ and $Data\ 3$, LSTM consistently outperform DNN, which confirms that LSTM is a better model than DNN when dealing with time series and sequence data, as expected. 

Figure~\ref{fig:lstm} is a sample plot of the LSTM predicted results vs ground truth during testing. It is observed that the predicted results (third plot in Figure~\ref{fig:lstm}) are very close to the ground truth (second plot in Figure~\ref{fig:lstm}), which confirmed the superior performance of LSTM in predicting time-series data. The convergence curve of the training and validation process of LSTM is given in Figure~\ref{fig:loss_lstm}. It can be seen that the loss decreased dramatically during the first 25 epochs and stayed low afterwards. 

\section{Related Research}
\label{sec:relatedwork}
In the time past, various researches bothering around channel modeling and the application of machine learning have been done to solve various problems. Transmission losses between transceivers, effects of bit error rate, maximum internode distances for different networks and depths, effect of weather season, and variability of ocean environmental factors were investigated in~\cite{chanmodelUWCN2008} through simulations based on ray-theory-based multipath Rayleigh underwater channel models for shallow and deep waters. In~\cite{chanmodelIOT2021}, the authors derived the channel path loss, modified the log-distance model to create a model suitable for an underwater IoT network  and designed an empirical channel model for the medium distance underwater acoustic channels based on real measurement data.

The underwater acoustic channel was modeled using the BELLHOP ray model in~\cite{MAUACCB2017} in order to analyze sound propagation characteristics putting into consideration the rough nature of the sea surfaces and the bottoms for different oceanic conditions. In~\cite{MACUWCS2012}, modeling the underwater acoustic channel was derived to mathematically quantify the channel characteristics for autonomous underwater vehicle (AUV) wireless communication system. The model was derived from the AN product, SNR and Band Selection where $A$ is the attenuation model, $N$ is the ambient noise model and SNR is the signal-to-noise ratio.

A deep learning network based signal detection was used for full-duplex cognitive underwater acoustic communication with self interface cancellation(SIC) as experimented in~\cite{SDFCUACMDLM2019}. Likewise, the combination of the convolutional neural network (CNN) and long short-term memory (LSTM) network was used in~\cite{ModClassUACDL2018} for the automatic modulation classification of underwater communication signals using dataset obtained from a modeled multi-path fading underwater acoustic channel with alpha-stable impulse noise and Doppler frequency shift.~\cite{BEAMCUAS2018} also experimented the use of blind equalization in conjunction with convolutional neural network (CNN) to lessen the effects of underwater channel limited bandwidth, multi-path, clutter,etc and to characterize acoustic signals automatically, in the automatic modulation classification of underwater acoustic signals. 

Youwen Zhang, $et$ $al.$ used the deep neural network (DNN) to build a deep learning based receiver in~\cite{DLBSCCUAC2019} for single carrier communication in an underwater acoustic channel using data gotten from the sea. The deep neural network (DNN) based receiver consistently performed better in different simulation configurations using features extracted by the deep network compared to the traditional channel-estimate (CE)-based decision feedback equalizer (DFE).~\cite{DNNChanEstUACOFDMRongkun2019} also used the DNN for channel estimation using data gotten from the Bellhop Ray model simulation of the underwater acoustic environment. When the experimental results were compared with those from the conventional channel estimation methods such as least square (LS), back propagation neural network (BPNN) and  minimum mean square error (MMSE), the DNN outperformed both the LS and BPNN algorithms and are only comparable to the MMSE algorithm in terms of the bit error rate and normalized mean square error.

By representing the receiver as a DNN in~\cite{ZHANG201953} and~\cite{UndAcouComDeepLearnYouwen2018}, a deep learning based underwater acoustic (UWA) orthogonal frequency-division multiplexing (OFDM) communication system was built. Without explicit channel estimation and equalization used, the deep learning UWA communication systems could directly recover the transmitted symbols after training. When the performance of the receiver was evaluated for various criteria such as the raining pilots and the size of training set, simulation results revealed that the deep learning based UWA OFDM communication is more robust. Gang Hu, et al developed a depth learning-based underwater target recognition method in~\cite{deepUnderWater2018} for underwater acoustic target classification and recognition using CNN and extreme learning machine (ELM). Though convolution neural networks can perform both feature extraction and classification, they rely on a full connection layer that is trained using gradient descent; thus their generalization capacity is limited and sub-optimal. Hence Gang used an extreme learning machine (ELM) in the classification stage. When the classifier used was compared to the traditional CNN classifier, the recognition rate was greatly improved.

An MLP network, a two connected multi-layer perceptron, was integrated into a receiver in~\cite{DLSDUOWC2020} for signal detection for single photon avalanche photodiode (SPAD)-based underwater optical wireless communication. Experiments were carried out in different water types and when the results were compared with other baselines, the MLP network's overall bit error ratio (BER) was better. The simulation results further confirmed that the deep learning method addressed the optical channel distortion, non-Poisson distortion and consequently improved the system performance. In~\cite{LAUCNML2020}, several machine learning methods were used to classify the modulation type in their bid to find an efficient link adaptation method depending on the channel quality on an underwater communications network because it was difficult to identify modulation during actual communication owing to the complex and unstable nature of the underwater acoustic communication systems. Other methods like the sparse adaptive convolution cores, time-domain turbo equalization and frequency-domain turbo equalization when used had the problem of high computational complexity and low classification success rate. The boosted regression tree showed accuracy of 99.97\% in classifying the modulation and coding scheme (MCS) levels out of the four different ML algorithms adopted for this experiment. 

\section{Conclusions}
\label{sec:conclusions}
In this paper, we explored the capability of deep learning and some traditional machine learning methods, to learn and accurately model the underwater acoustic channel using real underwater data collected from a water tank with disturbance and from a lake. Specifically we used the Deep Neural Network and the Long Short Term Memory to model the underwater acoustic channel. Experimental results show that deep learning can be used for modeling the underwater channel with better performance when compared to the traditional machine learning methods in terms of mean absolute percentage error. This work is unique in the sense that it uses real underwater data in modeling the channel and not based on some mathematical assumptions and approximations.

While deep learning demonstrated a good performance in modeling underwater acoustic communications, it requires large amount of data and intensive training. For instance, we used NVIDIA GPU servers to train the DNN and LSTM models in this work. Although the deep learning models gave a better result when compared to that of the traditional machine learning methods, it is still desirable to study other alternative methods that may provide even better performance with less computational complexity. We plan to explore Reservoir Computing in our future works. Reservoir computing has demonstrated to be a valuable tool for modeling and predicting dynamical systems using time-series data with less demanding computation~\cite{phoneme2010},~\cite{infoProcessing2011}. It was also demonstrated in~\cite{modelFree2018} that reservoir computing is capable of forecasting massive chaotic systems. Future works include investigating reservoir computing and how it can be applied to modeling the underwater acoustic channel for much better performance and less computational cost.
%%
%% The acknowledgments section is defined using the "acks" environment
%% (and NOT an unnumbered section). This ensures the proper
%% identification of the section in the article metadata, and the
%% consistent spelling of the heading.
%
\begin{acks}
\label{sec:acknowledgement}
We acknowledge the data source from Dr. Hao Xu's group at University of Nevada, Reno and helpful discussions with Dr. Hao Xu.
This research work is supported in part by the U.S. Dept. of Navy under agreement number N00014-17-1-3062 and the U.S. Office of the Under Secretary of Defense for Research and Engineering (OUSD(R\&E)) under agreement number FA8750-15-2-0119. The U.S. Government is authorized to reproduce and distribute reprints for governmental purposes notwithstanding any copyright notation thereon. The views and conclusions contained herein are those of the authors and should not be interpreted as necessarily representing the official policies or endorsements, either expressed or implied, of the Dept. of Navy  or the Office of the Under Secretary of Defense for Research and Engineering (OUSD(R\&E)) or the U.S. Government.
\end{acks}

%%
%% The next two lines define the bibliography style to be used, and
%% the bibliography file.
\bibliographystyle{ACM-Reference-Format}
\bibliography{UnderWaterChannelModel}

%%% -*-BibTeX-*-
%%% Do NOT edit. File created by BibTeX with style
%%% ACM-Reference-Format-Journals [18-Jan-2012].

\begin{thebibliography}{28}

%%% ====================================================================
%%% NOTE TO THE USER: you can override these defaults by providing
%%% customized versions of any of these macros before the \bibliography
%%% command.  Each of them MUST provide its own final punctuation,
%%% except for \shownote{}, \showDOI{}, and \showURL{}.  The latter two
%%% do not use final punctuation, in order to avoid confusing it with
%%% the Web address.
%%%
%%% To suppress output of a particular field, define its macro to expand
%%% to an empty string, or better, \unskip, like this:
%%%
%%% \newcommand{\showDOI}[1]{\unskip}   % LaTeX syntax
%%%
%%% \def \showDOI #1{\unskip}           % plain TeX syntax
%%%
%%% ====================================================================

\ifx \showCODEN    \undefined \def \showCODEN     #1{\unskip}     \fi
\ifx \showDOI      \undefined \def \showDOI       #1{#1}\fi
\ifx \showISBNx    \undefined \def \showISBNx     #1{\unskip}     \fi
\ifx \showISBNxiii \undefined \def \showISBNxiii  #1{\unskip}     \fi
\ifx \showISSN     \undefined \def \showISSN      #1{\unskip}     \fi
\ifx \showLCCN     \undefined \def \showLCCN      #1{\unskip}     \fi
\ifx \shownote     \undefined \def \shownote      #1{#1}          \fi
\ifx \showarticletitle \undefined \def \showarticletitle #1{#1}   \fi
\ifx \showURL      \undefined \def \showURL       {\relax}        \fi
% The following commands are used for tagged output and should be
% invisible to TeX
\providecommand\bibfield[2]{#2}
\providecommand\bibinfo[2]{#2}
\providecommand\natexlab[1]{#1}
\providecommand\showeprint[2][]{arXiv:#2}

\bibitem[\protect\citeauthoryear{Ahmed, Atiya, Gayar, and El-Shishiny}{Ahmed
  et~al\mbox{.}}{2010}]%
        {ECMLTS2010}
\bibfield{author}{\bibinfo{person}{Nesreen Ahmed}, \bibinfo{person}{Amir
  Atiya}, \bibinfo{person}{Neamat Gayar}, {and} \bibinfo{person}{Hisham
  El-Shishiny}.} \bibinfo{year}{2010}\natexlab{}.
\newblock \showarticletitle{An Empirical Comparison of Machine Learning Models
  for Time Series Forecasting}.
\newblock \bibinfo{journal}{\emph{Econometric Reviews}}  \bibinfo{volume}{29}
  (\bibinfo{date}{08} \bibinfo{year}{2010}), \bibinfo{pages}{594--621}.
\newblock
\urldef\tempurl%
\url{https://doi.org/10.1080/07474938.2010.481556}
\showDOI{\tempurl}


\bibitem[\protect\citeauthoryear{Alamgir, {Sultana}, and {Chang}}{Alamgir
  et~al\mbox{.}}{2020}]%
        {LAUCNML2020}
\bibfield{author}{\bibinfo{person}{M.S.M. Alamgir}, \bibinfo{person}{M.~N.
  {Sultana}}, {and} \bibinfo{person}{K. {Chang}}.}
  \bibinfo{year}{2020}\natexlab{}.
\newblock \showarticletitle{Link Adaptation on an Underwater Communications
  Network Using Machine Learning Algorithms: Boosted Regression Tree Approach}.
\newblock \bibinfo{journal}{\emph{IEEE Access}}  \bibinfo{volume}{8}
  (\bibinfo{year}{2020}), \bibinfo{pages}{73957--73971}.
\newblock


\bibitem[\protect\citeauthoryear{Appeltant, Soriano, Van~der Sande, Danckaert,
  Massar, Dambre, Schrauwen, Mirasso, and Fischer}{Appeltant
  et~al\mbox{.}}{2011}]%
        {infoProcessing2011}
\bibfield{author}{\bibinfo{person}{Lennert Appeltant}, \bibinfo{person}{Miguel
  Soriano}, \bibinfo{person}{Guy Van~der Sande}, \bibinfo{person}{Jan
  Danckaert}, \bibinfo{person}{Serge Massar}, \bibinfo{person}{J. Dambre},
  \bibinfo{person}{Benjamin Schrauwen}, \bibinfo{person}{Claudio Mirasso},
  {and} \bibinfo{person}{Ingo Fischer}.} \bibinfo{year}{2011}\natexlab{}.
\newblock \showarticletitle{Information processing using a single dynamical
  node as complex system}.
\newblock \bibinfo{journal}{\emph{Nature communications}}  \bibinfo{volume}{2}
  (\bibinfo{date}{09} \bibinfo{year}{2011}), \bibinfo{pages}{468}.
\newblock
\urldef\tempurl%
\url{https://doi.org/10.1038/ncomms1476}
\showDOI{\tempurl}


\bibitem[\protect\citeauthoryear{Bengio}{Bengio}{2009}]%
        {BengioNOW2009}
\bibfield{author}{\bibinfo{person}{Yoshua Bengio}.}
  \bibinfo{year}{2009}\natexlab{}.
\newblock \showarticletitle{Learning Deep Architectures for AI}.
\newblock \bibinfo{journal}{\emph{Foundations and Trends® in Machine
  Learning}} \bibinfo{volume}{2}, \bibinfo{number}{1} (\bibinfo{year}{2009}),
  \bibinfo{pages}{1--127}.
\newblock
\showISSN{1935-8237}
\urldef\tempurl%
\url{https://doi.org/10.1561/2200000006}
\showDOI{\tempurl}


\bibitem[\protect\citeauthoryear{{Chitre}}{{Chitre}}{2007}]%
        {ChitreHighFreq2007}
\bibfield{author}{\bibinfo{person}{M. {Chitre}}.}
  \bibinfo{year}{2007}\natexlab{}.
\newblock \showarticletitle{A high-frequency warm shallow water acoustic
  communications channel model and measurements}.
\newblock \bibinfo{journal}{\emph{The Journal of the Acoustical Society of
  America}} \bibinfo{volume}{122}, \bibinfo{number}{5} (\bibinfo{year}{2007}),
  \bibinfo{pages}{2580‐2586}.
\newblock


\bibitem[\protect\citeauthoryear{Domingo}{Domingo}{2008}]%
        {chanmodelUWCN2008}
\bibfield{author}{\bibinfo{person}{Mari~Carmen Domingo}.}
  \bibinfo{year}{2008}\natexlab{}.
\newblock \showarticletitle{Overview of Channel Models for Underwater Wireless
  Communication Networks}.
\newblock \bibinfo{journal}{\emph{Phys. Commun.}} \bibinfo{volume}{1},
  \bibinfo{number}{3} (\bibinfo{date}{Sept.} \bibinfo{year}{2008}),
  \bibinfo{pages}{163–182}.
\newblock
\showISSN{1874-4907}
\urldef\tempurl%
\url{https://doi.org/10.1016/j.phycom.2008.09.001}
\showDOI{\tempurl}


\bibitem[\protect\citeauthoryear{Gang, Wang, Peng, Qiu, Shi, and Liu}{Gang
  et~al\mbox{.}}{2018}]%
        {deepUnderWater2018}
\bibfield{author}{\bibinfo{person}{Hu Gang}, \bibinfo{person}{Kejun Wang},
  \bibinfo{person}{Yuan Peng}, \bibinfo{person}{Mengran Qiu},
  \bibinfo{person}{Jianfei Shi}, {and} \bibinfo{person}{Liangliang Liu}.}
  \bibinfo{year}{2018}\natexlab{}.
\newblock \showarticletitle{Deep Learning Methods for Underwater Target Feature
  Extraction and Recognition}.
\newblock \bibinfo{journal}{\emph{Computational Intelligence and Neuroscience}}
   \bibinfo{volume}{2018} (\bibinfo{date}{03} \bibinfo{year}{2018}),
  \bibinfo{pages}{1--10}.
\newblock
\urldef\tempurl%
\url{https://doi.org/10.1155/2018/1214301}
\showDOI{\tempurl}


\bibitem[\protect\citeauthoryear{Goodfellow, Bengio, and Courville}{Goodfellow
  et~al\mbox{.}}{2016}]%
        {GoodBengCour2016}
\bibfield{author}{\bibinfo{person}{Ian~J. Goodfellow}, \bibinfo{person}{Yoshua
  Bengio}, {and} \bibinfo{person}{Aaron Courville}.}
  \bibinfo{year}{2016}\natexlab{}.
\newblock \bibinfo{booktitle}{\emph{Deep Learning}}.
\newblock \bibinfo{publisher}{MIT Press}, \bibinfo{address}{Cambridge, MA,
  USA}.
\newblock
\newblock
\shownote{\url{http://www.deeplearningbook.org}}.


\bibitem[\protect\citeauthoryear{Hochreiter and Schmidhuber}{Hochreiter and
  Schmidhuber}{1997}]%
        {HochreiterACM1997}
\bibfield{author}{\bibinfo{person}{Sepp Hochreiter} {and}
  \bibinfo{person}{J\"{u}rgen Schmidhuber}.} \bibinfo{year}{1997}\natexlab{}.
\newblock \showarticletitle{Long Short-Term Memory}.
\newblock \bibinfo{journal}{\emph{Neural Comput.}} \bibinfo{volume}{9},
  \bibinfo{number}{8} (\bibinfo{date}{Nov.} \bibinfo{year}{1997}),
  \bibinfo{pages}{1735--1780}.
\newblock
\showISSN{0899-7667}
\urldef\tempurl%
\url{https://doi.org/10.1162/neco.1997.9.8.1735}
\showDOI{\tempurl}


\bibitem[\protect\citeauthoryear{{Jiang}, {Cao}, {Xue}, and {Tang}}{{Jiang}
  et~al\mbox{.}}{2017}]%
        {MAUACCB2017}
\bibfield{author}{\bibinfo{person}{R. {Jiang}}, \bibinfo{person}{S. {Cao}},
  \bibinfo{person}{C. {Xue}}, {and} \bibinfo{person}{L. {Tang}}.}
  \bibinfo{year}{2017}\natexlab{}.
\newblock \showarticletitle{Modeling and analyzing of underwater acoustic
  channels with curvilinear boundaries in shallow ocean}. In
  \bibinfo{booktitle}{\emph{2017 IEEE International Conference on Signal
  Processing, Communications and Computing (ICSPCC)}}. \bibinfo{pages}{1--6}.
\newblock


\bibitem[\protect\citeauthoryear{{Jiang}, {Sun}, {Zhang}, {Tang}, {Wang}, and
  {Zhang}}{{Jiang} et~al\mbox{.}}{2020}]%
        {DLSDUOWC2020}
\bibfield{author}{\bibinfo{person}{R. {Jiang}}, \bibinfo{person}{C. {Sun}},
  \bibinfo{person}{L. {Zhang}}, \bibinfo{person}{X. {Tang}},
  \bibinfo{person}{H. {Wang}}, {and} \bibinfo{person}{A. {Zhang}}.}
  \bibinfo{year}{2020}\natexlab{}.
\newblock \showarticletitle{Deep Learning Aided Signal Detection for SPAD-Based
  Underwater Optical Wireless Communications}.
\newblock \bibinfo{journal}{\emph{IEEE Access}}  \bibinfo{volume}{8}
  (\bibinfo{year}{2020}), \bibinfo{pages}{20363--20374}.
\newblock


\bibitem[\protect\citeauthoryear{{Jiang}, {Wang}, {Cao}, {Zhao}, and
  {Li}}{{Jiang} et~al\mbox{.}}{2019}]%
        {DNNChanEstUACOFDMRongkun2019}
\bibfield{author}{\bibinfo{person}{R. {Jiang}}, \bibinfo{person}{X. {Wang}},
  \bibinfo{person}{S. {Cao}}, \bibinfo{person}{J. {Zhao}}, {and}
  \bibinfo{person}{X. {Li}}.} \bibinfo{year}{2019}\natexlab{}.
\newblock \showarticletitle{Deep Neural Networks for Channel Estimation in
  Underwater Acoustic OFDM Systems}.
\newblock \bibinfo{journal}{\emph{IEEE Access}}  \bibinfo{volume}{7}
  (\bibinfo{year}{2019}), \bibinfo{pages}{23579--23594}.
\newblock
\showISSN{2169-3536}
\urldef\tempurl%
\url{https://doi.org/10.1109/ACCESS.2019.2899990}
\showDOI{\tempurl}


\bibitem[\protect\citeauthoryear{Le, Ho, Lee, and Jung}{Le
  et~al\mbox{.}}{2019}]%
        {applstm2019}
\bibfield{author}{\bibinfo{person}{Xuan~Hien Le}, \bibinfo{person}{Hung Ho},
  \bibinfo{person}{Giha Lee}, {and} \bibinfo{person}{Sungho Jung}.}
  \bibinfo{year}{2019}\natexlab{}.
\newblock \showarticletitle{Application of Long Short-Term Memory (LSTM) Neural
  Network for Flood Forecasting}.
\newblock \bibinfo{journal}{\emph{Water}}  \bibinfo{volume}{11}
  (\bibinfo{date}{07} \bibinfo{year}{2019}), \bibinfo{pages}{1387}.
\newblock
\urldef\tempurl%
\url{https://doi.org/10.3390/w11071387}
\showDOI{\tempurl}


\bibitem[\protect\citeauthoryear{LeCun, Bengio, and Hinton}{LeCun
  et~al\mbox{.}}{2015}]%
        {LeCunNature2015}
\bibfield{author}{\bibinfo{person}{Yann LeCun}, \bibinfo{person}{Yoshua
  Bengio}, {and} \bibinfo{person}{Geoffrey Hinton}.}
  \bibinfo{year}{2015}\natexlab{}.
\newblock \showarticletitle{{Deep learning}}.
\newblock \bibinfo{journal}{\emph{Nature}} \bibinfo{volume}{521},
  \bibinfo{number}{7553} (\bibinfo{date}{27 May} \bibinfo{year}{2015}),
  \bibinfo{pages}{436--444}.
\newblock
\showISSN{0028-0836}
\urldef\tempurl%
\url{https://doi.org/10.1038/nature14539}
\showDOI{\tempurl}


\bibitem[\protect\citeauthoryear{Lee and Lee}{Lee and Lee}{2021}]%
        {chanmodelIOT2021}
\bibfield{author}{\bibinfo{person}{Ho-Kyoung Lee} {and} \bibinfo{person}{Byung
  Lee}.} \bibinfo{year}{2021}\natexlab{}.
\newblock \showarticletitle{An Underwater Acoustic Channel Modeling for
  Internet of Things Networks}.
\newblock \bibinfo{journal}{\emph{Wireless Personal Communications}}
  \bibinfo{volume}{116} (\bibinfo{date}{02} \bibinfo{year}{2021}).
\newblock
\urldef\tempurl%
\url{https://doi.org/10.1007/s11277-020-07817-x}
\showDOI{\tempurl}


\bibitem[\protect\citeauthoryear{{Li-Da}, {Shi-Lian}, and {Wei}}{{Li-Da}
  et~al\mbox{.}}{2018}]%
        {ModClassUACDL2018}
\bibfield{author}{\bibinfo{person}{D. {Li-Da}}, \bibinfo{person}{W.
  {Shi-Lian}}, {and} \bibinfo{person}{Z. {Wei}}.}
  \bibinfo{year}{2018}\natexlab{}.
\newblock \showarticletitle{Modulation Classification of Underwater Acoustic
  Communication Signals Based on Deep Learning}. In
  \bibinfo{booktitle}{\emph{2018 OCEANS - MTS/IEEE Kobe Techno-Oceans (OTO)}}.
  \bibinfo{pages}{1--4}.
\newblock


\bibitem[\protect\citeauthoryear{Luo, Pu, Zuba, Peng, and Cui}{Luo
  et~al\mbox{.}}{2014}]%
        {LuoCognitive2014}
\bibfield{author}{\bibinfo{person}{Yu Luo}, \bibinfo{person}{Lina Pu},
  \bibinfo{person}{Michael Zuba}, \bibinfo{person}{Zheng Peng}, {and}
  \bibinfo{person}{Jun-Hong Cui}.} \bibinfo{year}{2014}\natexlab{}.
\newblock \showarticletitle{Cognitive acoustics: making underwater
  communications environment-friendly}. In
  \bibinfo{booktitle}{\emph{Conference: In Proceedings of the International
  Conference on Underwater Networks \& Systems (WUWNet)}}.
\newblock


\bibitem[\protect\citeauthoryear{Marcoux, Chandna, Egnor, and Blair}{Marcoux
  et~al\mbox{.}}{2018}]%
        {BEAMCUAS2018}
\bibfield{author}{\bibinfo{person}{Caitlyn~N. Marcoux}, \bibinfo{person}{Bindu
  Chandna}, \bibinfo{person}{Dianne Egnor}, {and} \bibinfo{person}{Ballard
  Blair}.} \bibinfo{year}{2018}\natexlab{}.
\newblock \showarticletitle{Blind equalization and automatic modulation
  classification of underwater acoustic signals}.
\newblock \bibinfo{journal}{\emph{Proceedings of Meetings on Acoustics}}
  \bibinfo{volume}{35}, \bibinfo{number}{1} (\bibinfo{year}{2018}),
  \bibinfo{pages}{055003}.
\newblock
\urldef\tempurl%
\url{https://doi.org/10.1121/2.0000952}
\showDOI{\tempurl}
\showeprint{https://asa.scitation.org/doi/pdf/10.1121/2.0000952}


\bibitem[\protect\citeauthoryear{Morozs, Gorma, Henson, Shen, Mitchell, and
  Zakharov}{Morozs et~al\mbox{.}}{2020}]%
        {chanmodelUAN2020}
\bibfield{author}{\bibinfo{person}{Nils Morozs}, \bibinfo{person}{Wael Gorma},
  \bibinfo{person}{Benjamin~T. Henson}, \bibinfo{person}{Lu Shen},
  \bibinfo{person}{Paul~D. Mitchell}, {and} \bibinfo{person}{Yuriy~V.
  Zakharov}.} \bibinfo{year}{2020}\natexlab{}.
\newblock \showarticletitle{Channel Modeling for Underwater Acoustic Network
  Simulation}.
\newblock \bibinfo{journal}{\emph{IEEE Access}}  \bibinfo{volume}{8}
  (\bibinfo{year}{2020}), \bibinfo{pages}{136151--136175}.
\newblock
\showISSN{2169-3536}
\urldef\tempurl%
\url{https://doi.org/10.1109/ACCESS.2020.3011620}
\showDOI{\tempurl}


\bibitem[\protect\citeauthoryear{Pathak, Hunt, Girvan, Lu, and Ott}{Pathak
  et~al\mbox{.}}{2018}]%
        {modelFree2018}
\bibfield{author}{\bibinfo{person}{Jaideep Pathak}, \bibinfo{person}{Brian
  Hunt}, \bibinfo{person}{Michelle Girvan}, \bibinfo{person}{Zhixin Lu}, {and}
  \bibinfo{person}{Edward Ott}.} \bibinfo{year}{2018}\natexlab{}.
\newblock \showarticletitle{Model-Free Prediction of Large Spatiotemporally
  Chaotic Systems from Data: A Reservoir Computing Approach}.
\newblock \bibinfo{journal}{\emph{Physical Review Letters}}
  \bibinfo{volume}{120} (\bibinfo{date}{01} \bibinfo{year}{2018}).
\newblock
\urldef\tempurl%
\url{https://doi.org/10.1103/PhysRevLett.120.024102}
\showDOI{\tempurl}


\bibitem[\protect\citeauthoryear{Qian, Zhu, and Zhang}{Qian
  et~al\mbox{.}}{2017}]%
        {QianJCIN2017}
\bibfield{author}{\bibinfo{person}{Lijun Qian}, \bibinfo{person}{Jinkang Zhu},
  {and} \bibinfo{person}{Sihai Zhang}.} \bibinfo{year}{2017}\natexlab{}.
\newblock \showarticletitle{Survey of wireless big data}.
\newblock \bibinfo{journal}{\emph{Journal of Communications and Information
  Networks}} \bibinfo{volume}{2}, \bibinfo{number}{1} (\bibinfo{year}{2017}),
  \bibinfo{pages}{1--18}.
\newblock
\showISSN{2509-3312}
\urldef\tempurl%
\url{https://doi.org/10.1007/s41650-017-0001-2}
\showDOI{\tempurl}


\bibitem[\protect\citeauthoryear{{Stojanovic} and {Preisig}}{{Stojanovic} and
  {Preisig}}{2009}]%
        {StojanovicUWA2009}
\bibfield{author}{\bibinfo{person}{M. {Stojanovic}} {and} \bibinfo{person}{J.
  {Preisig}}.} \bibinfo{year}{2009}\natexlab{}.
\newblock \showarticletitle{Underwater acoustic communication channels:
  Propagation models and statistical characterization}.
\newblock \bibinfo{journal}{\emph{IEEE Communications Magazine}}
  \bibinfo{volume}{47}, \bibinfo{number}{1} (\bibinfo{year}{2009}),
  \bibinfo{pages}{84--89}.
\newblock


\bibitem[\protect\citeauthoryear{Triefenbach, Jalalvand, Schrauwen, and
  Martens}{Triefenbach et~al\mbox{.}}{2010}]%
        {phoneme2010}
\bibfield{author}{\bibinfo{person}{Fabian Triefenbach},
  \bibinfo{person}{Azarakhsh Jalalvand}, \bibinfo{person}{Benjamin Schrauwen},
  {and} \bibinfo{person}{jean-pierre Martens}.}
  \bibinfo{year}{2010}\natexlab{}.
\newblock \showarticletitle{Phoneme Recognition with Large Hierarchical
  Reservoirs.}. In \bibinfo{booktitle}{\emph{Advances in Neural Information
  Processing Systems 23: 24th Annual Conference on Neural Information
  Processing Systems 2010}}. \bibinfo{pages}{2307--2315}.
\newblock


\bibitem[\protect\citeauthoryear{{Wang}, {Ma}, {Cui}, {Sun}, {Zhou}, {Wang},
  {Li}, and {Liu}}{{Wang} et~al\mbox{.}}{2019}]%
        {SDFCUACMDLM2019}
\bibfield{author}{\bibinfo{person}{J. {Wang}}, \bibinfo{person}{S. {Ma}},
  \bibinfo{person}{Y. {Cui}}, \bibinfo{person}{H. {Sun}}, \bibinfo{person}{M.
  {Zhou}}, \bibinfo{person}{B. {Wang}}, \bibinfo{person}{J. {Li}}, {and}
  \bibinfo{person}{L. {Liu}}.} \bibinfo{year}{2019}\natexlab{}.
\newblock \showarticletitle{Signal Detection for Full-duplex Cognitive
  Underwater Acoustic Communications with SIC Using Model-Driven Deep Learning
  Network}. In \bibinfo{booktitle}{\emph{2019 IEEE International Conference on
  Signal Processing, Communications and Computing (ICSPCC)}}.
  \bibinfo{pages}{1--6}.
\newblock


\bibitem[\protect\citeauthoryear{{Yoong}, {Yeo}, {Teo}, and {Wong}}{{Yoong}
  et~al\mbox{.}}{2012}]%
        {MACUWCS2012}
\bibfield{author}{\bibinfo{person}{H.~P. {Yoong}}, \bibinfo{person}{K.~B.
  {Yeo}}, \bibinfo{person}{K.~T.~K. {Teo}}, {and} \bibinfo{person}{W.~L.
  {Wong}}.} \bibinfo{year}{2012}\natexlab{}.
\newblock \showarticletitle{Modeling of Acoustic Channel for Underwater
  Wireless Communication System in AUV Application}. In
  \bibinfo{booktitle}{\emph{2012 UKSim 14th International Conference on
  Computer Modelling and Simulation}}. \bibinfo{pages}{603--607}.
\newblock


\bibitem[\protect\citeauthoryear{Zhang, Li, Zakharov, Li, and Li}{Zhang
  et~al\mbox{.}}{2019}]%
        {ZHANG201953}
\bibfield{author}{\bibinfo{person}{Youwen Zhang}, \bibinfo{person}{Junxuan Li},
  \bibinfo{person}{Yuriy Zakharov}, \bibinfo{person}{Xiang Li}, {and}
  \bibinfo{person}{Jianghui Li}.} \bibinfo{year}{2019}\natexlab{}.
\newblock \showarticletitle{Deep learning based underwater acoustic OFDM
  communications}.
\newblock \bibinfo{journal}{\emph{Applied Acoustics}}  \bibinfo{volume}{154}
  (\bibinfo{year}{2019}), \bibinfo{pages}{53 -- 58}.
\newblock
\showISSN{0003-682X}
\urldef\tempurl%
\url{https://doi.org/10.1016/j.apacoust.2019.04.023}
\showDOI{\tempurl}


\bibitem[\protect\citeauthoryear{Zhang, Li, Zakharov, Sun, and Li}{Zhang
  et~al\mbox{.}}{2018}]%
        {UndAcouComDeepLearnYouwen2018}
\bibfield{author}{\bibinfo{person}{Youwen Zhang}, \bibinfo{person}{Junxuan Li},
  \bibinfo{person}{Yuriy Zakharov}, \bibinfo{person}{Dajun Sun}, {and}
  \bibinfo{person}{Jianghui Li}.} \bibinfo{year}{2018}\natexlab{}.
\newblock \showarticletitle{Underwater acoustic OFDM communications using deep
  learning}. In \bibinfo{booktitle}{\emph{The 2nd Franco-Chinese Acoustic
  Conference (FCAC) (31/10/18)}}.
\newblock
\urldef\tempurl%
\url{https://eprints.soton.ac.uk/426097/}
\showURL{%
\tempurl}


\bibitem[\protect\citeauthoryear{{Zhang}, {Li}, {Zakharov}, {Li}, {Li}, {Lin},
  and {Li}}{{Zhang} et~al\mbox{.}}{2019}]%
        {DLBSCCUAC2019}
\bibfield{author}{\bibinfo{person}{Y. {Zhang}}, \bibinfo{person}{J. {Li}},
  \bibinfo{person}{Y.~V. {Zakharov}}, \bibinfo{person}{J. {Li}},
  \bibinfo{person}{Y. {Li}}, \bibinfo{person}{C. {Lin}}, {and}
  \bibinfo{person}{X. {Li}}.} \bibinfo{year}{2019}\natexlab{}.
\newblock \showarticletitle{Deep Learning Based Single Carrier Communications
  Over Time-Varying Underwater Acoustic Channel}.
\newblock \bibinfo{journal}{\emph{IEEE Access}}  \bibinfo{volume}{7}
  (\bibinfo{year}{2019}), \bibinfo{pages}{38420--38430}.
\newblock


\end{thebibliography}

\end{document}